\documentclass{article}

\usepackage{microtype}
\usepackage{graphicx}
\usepackage{subfigure}
\usepackage{booktabs} % for professional tables

\usepackage{hyperref}

\usepackage[accepted]{icml2021}
\usepackage{amsfonts}
\usepackage{amsmath}

\newtheorem{definition}{Definition}

\icmltitlerunning{Graph Reordering for Cache-Efficient Near Neighbor Search}

\begin{document}

\twocolumn[
\icmltitle{Graph Reordering for Cache-Efficient Near Neighbor Search}



\icmlsetsymbol{equal}{*}

\begin{icmlauthorlist}
\icmlauthor{Benjamin Coleman}{ece,equal}
\icmlauthor{Santiago Segarra}{ece}
\icmlauthor{Alex Smola}{amzn}
\icmlauthor{Anshumali Shrivastava}{ece,cs}
\end{icmlauthorlist}

\icmlaffiliation{ece}{Department of Electrical and Computer Engineering, Rice University, Houston, TX}
\icmlaffiliation{cs}{Department of Computer Science, Rice University, Houston, TX}
\icmlaffiliation{amzn}{Amazon Web Services}

\icmlcorrespondingauthor{Benjamin Coleman}{ben.coleman@rice.edu}
\icmlcorrespondingauthor{Anshumali Shrivastava}{anshumali@rice.edu}
\icmlcorrespondingauthor{Alex Smola}{alex@smola.org}

\icmlkeywords{Machine Learning, ICML}

\vskip 0.3in
]



\printAffiliationsAndNotice{\icmlEqualContribution} % otherwise use the standard text.

\begin{abstract}
Graph search is one of the most successful algorithmic trends in near neighbor search. Several of the most popular and empirically successful algorithms are, at their core, a simple walk along a pruned near neighbor graph. Such algorithms consistently perform at the top of industrial speed benchmarks for applications such as embedding search. However, graph traversal applications often suffer from poor memory access patterns, and near neighbor search is no exception to this rule. Our measurements show that popular search indices such as the hierarchical navigable small-world graph (HNSW) can have poor cache miss performance. To address this problem, we apply graph reordering algorithms to near neighbor graphs. Graph reordering is a memory layout optimization that groups commonly-accessed nodes together in memory. 
We present exhaustive experiments applying several reordering algorithms to a leading graph-based near neighbor method based on the HNSW index.
We find that reordering improves the query time by up to 40\%, and we demonstrate that the time needed to reorder the graph is negligible compared to the time required to construct the index.
\end{abstract}

\section{Introduction}

Near neighbor search is a fundamental building block within many applications in machine learning systems. 
Informally, the task can be understood as follows. 
Given a dataset $D = \{x_1, x_2, ... x_N\}$, we wish to build a \emph{data structure} that can be queried with any point $q$ to obtain the $k$ points $x_i \in D$ that have the smallest distance to the query. This data structure is called a \textit{near neighbor index}. Near neighbor indices are of tremendous practical importance, as they form the backbone of production models in recommendation systems, natural language processing~\cite{mikolov2013distributed}, genomics~\cite{ondov2016mash}, computer vision~\cite{lowe1999object} and other applications in machine learning~\cite{shakhnarovich2008nearest}.
% \blue{References?}
% Ben: Piotyr's book is a bit of an older reference, but applications are still relevant

\textbf{Applications.} Near neighbor search has always been the core of canonical learning algorithms such as neighbor-based classification and regression. However, the problem has recently become the focus of intense research activity due to its central role in the task of prediction with embedding models. In a neural embedding model, objects are transformed into \textit{embeddings} in $\mathbb{R}^d$, where $d$ ranges from 100 to 1000 and $N$ often exceeds 100 million. Prediction typically consists of finding the $k$ nearest embeddings. For example, Amazon's deep semantic search engine works by embedding the product catalog and the search query into the same high-dimensional space. The engine recommends the products whose embeddings are nearest to the embedded search query~\cite{nigam2019semantic}.

Since the search occurs for every query, the latency and recall of the machine learning system critically depend on the ability to perform fast near neighbor search in the high-recall regime. Similar problems exist in natural language processing~\cite{kusner2015word}, information retrieval~\cite{huang2013learning}, computer vision~\cite{frome2013devise} and other application domains. 
Beyond machine learning, near neighbor indices are used to match audio recordings, detect plagiarism, classify ECG signals in healthcare, and perform copyright attribution with blockchain~\cite{jafari2021survey}. As a result, near neighbor search has become a central bottleneck for many applications.
% \red{recommendation systems} \blue{but the example we gave was already about a recommendation system, right?},

\textbf{Latency Challenges.} Solutions to the near neighbor problem are incredibly diverse, ranging from hardware-accelerated brute force~\cite{johnson2019billion} to space-partitioning trees~\cite{beygelzimer2006cover}.
Graph algorithms have emerged as a markedly effective class of methods for high-dimensional near neighbor search. For example, four of the top five search libraries on the well-established ANN-benchmarks leader board use a graph index~\cite{aumuller2020ann}. Because these libraries have been integrated with many large-scale production systems~\cite{JDH17,malkov2018efficient}, they have been hand-optimized in C++ at a large engineering cost. Systems for fast near neighbor search are the focus of a highly active research area, and optimizations have a clear practical impact. 

In this paper, we investigate the possibility of using \textit{graph reordering} to improve approximate near neighbor search. Graph reordering is a cache optimization that works by placing neighboring nodes in consecutive (or near-consecutive) memory locations. When a node is loaded into memory by a graph processing algorithm, modern CPU architectures will automatically load nodes from nearby memory locations as well. By reordering the nodes, we ensure that nodes which are likely to be visited by the algorithm are already pre-loaded into the CPU cache. The relabeling process may seem simple, but RAM access is responsible for up to 70\% of the runtime for tasks such as PageRank, graph decomposition, and graph diameter computation~\cite{wei2016speedup}. 
Since cache hits are an order of magnitude faster than memory access, reordering can speed up graph applications by 50\% or more~\cite{balaji2018graph}. Our work reveals that near neighbor search can be significantly accelerated through graph reordering, a hitherto unexplored optimization for this important machine learning task. 

% \blue{the reference has no year?} (DONE)
% \red{Based on our cache miss experiments with popular libraries, we suggest that near neighbor search is also a prime candidate for graph reordering.}
% \blue{Maybe a bit stronger sentence, what about the following?}
% \green{Our work reveals that near neighbor search can also be significantly accelerated through graph reordering, a hitherto unexplored strategy on this popular machine learning task.}
% Ben: I quite like this, yes

\subsection{Our Contribution}

% We perform an exhaustive set of experiments on embedding \red{search tasks such as SIFT, GIST and DEEP1B vectors} \blue{This reads weird, you are naming the `task' as `SIFT vectors' ... maybe we can delete the word `vectors'. Also, some reference here might be good.}. 

We integrate six recent graph ordering algorithms into the hierarchical navigable small-world graph (HNSW), a leading graph index that provides state-of-the-art performance on common benchmarks. We perform an exhaustive comparison on the SIFT100M, GIST, and DEEP100M datasets where we benchmark the query time and cache miss rate. Our experiments show that graph reordering is a viable and effective way to improve practical machine learning deployments that involve near neighbor search. We make the following new observations:

% We benchmark the query time and profile the number of cache misses that occur during the search. We observe that reordering reduces the cache miss rate by up to 10\% \red{(NOTE: update with final results)} \blue{marking this in color so that we don't forget} and the query latency by up to 25\%. We compare reordering algorithms and find that recently-proposed \textit{lightweight} graph algorithms do not improve near neighbor search; the NP-hard formulation seems to be necessary. Fortunately, the \green{incurred} reordering cost is negligible compared to the $k$-NN index construction cost \blue{is this based on the heuristics approximating the NP-hard solution? It seems weird to say that something is NP-hard and then say that the cost of solving it is negligible}. 

% \blue{Maybe we can put the answers to these questions instead of the questions themselves?}
% Great suggestion - added to paper

\begin{itemize}
    \item Graph reordering substantially improves the query time of near neighbor search, with speedups of 10-40\% on large embedding datasets. 
    \item Objective-based techniques, which optimize the node order to maximize a global cache coherence score, have the best performance. Recently-proposed \textit{lightweight} graph algorithms, which only use local node features, do not improve near neighbor search. % Algorithms based on the degree distribution of the graph Lightweight algorithms based that are effective for other applications rarely improve the query time on near neighbor graphs. 
    % Is it feasible to compute the reordering function for large-scale machine learning applications?
    \item Although objective-based methods are among the most expensive reordering algorithms, the reordering time is often an order of magnitude smaller than the time needed to construct the search index. 
    % Which reordering techniques are most appropriate for near neighbor search in production environments?
\end{itemize}

The question of whether reordering can improve near neighbor search is nontrivial because reordering can actually \textit{slow down} graph processing, depending on the characteristics of the input graph and the node access pattern of the application~\cite{balaji2018graph}. It should be noted that near neighbor graphs have much different structural properties than the graphs that typically benefit from reordering. For example, reordering algorithms often obtain large speedups on graphs with power-law degree distributions~\cite{faldu2019closer}, which near neighbor graphs do not have. 
It is also not clear whether the process of reordering a search index is prohibitively expensive. 
For commonly adopted formulations of the reordering problem, finding the optimal node labeling function is an NP-hard problem~\cite{wei2016speedup}.
Approximate algorithms and heuristics exist, but their complexity scales quadratically with the sum of node degrees. 
This is not an issue for graphs like social networks, which have a low average degree by virtue of their power-law distribution, but could pose serious problems when reordering near neighbor graphs, which are more densely connected. For industrial-scale embedding search, the graph index is frequently rebuilt to reflect changes in the model. Therefore, it is critical that our performance improvements do not substantially inflate the index construction time. 

% Our experiments show that graph reordering is a viable and effective way to improve practical machine learning deployments that involve near neighbor search.
% runtime issues have posed serious problems for other applications.

\section{Background and Related Work}

% \subsection{CPU Cache Hierarchy}

% \subsection{Fast Implementations of Near Neighbor Search}

% Algorithms for graph-based near neighbor search have been extensively studied in the context of embedding search. Methods 

% \textbf{CPU Cache Hierarchy.} Memory access Modern processors use caches to \blue{something missing here ...}
% Ben: forgot to remove, not sure if necessary

\textbf{Near Neighbor Graphs.} Graph-based algorithms such as HNSW, pruned approximate near neighbor graphs (PANNG)~\cite{iwasaki2016pruned}, and optimized near neighbor graphs (ONNG)~\cite{iwasaki2018optimization} feature prominently in leader boards and production systems at web-scale companies. 
Apart from recent theoretical progress~\cite{prokhorenkova2020graph}, the primary focus of graph index research has been to develop heuristics (such as diversification, pruning, and hierarchical structures) that improve the properties of the graph, increase search accuracy, and reduce search time.
In Section~\ref{sec:graph_knn}, we show that most search algorithms have essentially the same computational workload because these heuristics produce highly similar graph structures.
% can be reduced to essentially the same graph search procedure with minor differences that do not substantially affect the computational workload.
% \blue{This sentence is confusing ... the heursitics (diversification, pruning) are not themselves a graph search procedure, so how can they reduce to the `same graph search procedure'?}

An alternative method to reduce the search time is to develop implementation techniques that can be used to build fast systems. Due to their industrial importance, search indices have been optimized aggressively. However, there has been surprisingly little work to improve the cache locality properties of graph-based near neighbor search, even though graph algorithms are known to have poor cache performance. \citet{boystov2013engineeringefficient} describe memory fragmentation as an important consideration for HNSW, and the HNSW software manual~\cite{naidan2015non} describes a ``flattened index'' to improve the memory access pattern of near neighbor search. The flat layout stores nodes and data together in a contiguous block of memory without using pointers. The efficient layout improves cache performance, but it is agnostic to the graph structure. Graph reordering is a compatible technique that uses information about the graph to further improve the layout.

\textbf{Ordering with Space-Filling Curves.} The closest idea related to graph reordering is likely the use of space-filling curves for search problems~\cite{chan2002closest}. A space-filling curve is a line that passes through each point in the dataset. By minimizing the length of this line, we obtain a point ordering that places nearby points in consecutive positions. 
This idea has mainly been used to implement cache-efficient near neighbor search via a linear scan through the (ordered) data but has also been applied to graphs. \citet{connor2010fast} use the Morton ordering (a specific kind of space-filling curve) to improve the cache performance of graph construction for 3D point clouds. However, this work does not apply to machine learning applications such as embedding search for two important reasons. First, space-filling curves suffer from the curse of dimensionality and are not an effective way to order the high-dimensional vectors commonly encountered in embedding search. Second, existing work uses space-partitioning curves and reordering to speed up the \textit{construction} of the graph, while we are interested in whether reordering can improve the \textit{query time}. To the best of our knowledge, this paper is the only investigation of graph reordering for fast near neighbor queries.
% \blue{when you are using the reference as a noun, use the command citet, just like I did here} 

\section{Core Components of a Graph Search Index}
\label{sec:graph_knn}
% Introduce graph NNS

In this section, we introduce the key ingredients behind graph-based search indexing. 
As will become apparent, popular graph-based indices (HNSW, PANNG, and ONNG) share the same core components. Thus, although we focus on HNSW for our experiments, we expect our findings to generalize to most graph-based indices used in practice.
From an implementation perspective, the core components of a graph index are \textit{diversification / pruning}, \textit{search initialization}, and \textit{beam search}. 
Although major search algorithms do differ in nontrivial ways, these differences do not affect the node access pattern of the application or the properties of the graph that are relevant to graph reordering.

% We argue that the differences between major search algorithms do not affect the properties of the graph that are relevant to the node access pattern and graph reordering. 

% We argue that When viewed in this conceptual framework, most graph indices have the same node access pattern and perform the same search process over graphs that are highly similar.

% perform the same search algorithm over essentially the same graph.
% We also characterize the performance impact of search initialization and beam search.

In a near neighbor graph index, each node corresponds to an element from the dataset. Two nodes are connected if there is a small distance between the corresponding elements. To the best of our knowledge, all existing theoretical models of near neighbor graphs adopt one of the following two construction processes~\cite{laarhoven2017graph,sebastian2002metric}, where we assume the existence of a metric $d: V \to V$ between every pair of nodes.

% \blue{Notice that I deleted the `title' in the following definitions ... they are too short to merit a title in my view. Also, I deleted the notation $G(V,E)$, since you were not using it.}

\begin{definition}\normalfont
A graph is an \emph{$\alpha$-near neighbor ($\alpha$-NN)} graph if there is an edge between all nodes $v_1$ and $v_2$ such that $d(v_1,v_2) < \alpha$.
\end{definition}

\begin{definition}\normalfont 
A graph is a \emph{$k$-nearest neighbor ($k$-NN)} graph if each node $v$ is connected to its $k$ nearest nodes.
\end{definition}

The two graph models differ by degree distribution and connectivity. In the $k$-NN graph, each node has an out-degree equal to $k$, while nodes in the $\alpha$-NN graph may be connected to a variable number of neighbors. 
Furthermore, $k$-NN graphs are directed, while $\alpha$-NN graphs are undirected because the distance function $d$ is symmetric. 
It should be noted that both models may produce disconnected graphs, with nodes in sparsely populated regions of the dataset forming separate connected components.

\subsection{Constructing Near Neighbor Graphs}

In practice, one cannot efficiently construct the $k$-NN or $\alpha$-NN graphs because $O(N^2)$ computations are required to find the nearest neighbors of each point in the dataset. 
Nearly all modern algorithms solve this chicken-and-egg problem via ``bootstrapping,'' where we iteratively add nodes to a partially-constructed graph. 
To add a node $q$ to the graph, we query the current version of the graph to find the (approximate) neighbors of $q$, whose edges are then updated to include $q$.
The result is an approximate $k$-NN or $\alpha$-NN graph. An alternative to search-based construction is to begin with a randomly-initialized graph and apply a procedure called \textit{NN-Descent}, where we query and update each node in the graph~\cite{dong2011efficient}.
Because the technique converges to a close approximation of the true graph in a small number of iterations, one may also use it to \textit{refine} a graph obtained using other methods.
% NN-Descent empirically converges to a close approximation of the true graph in a small number of iterations.
% using the same procedure, by querying and updating each node in the graph. 
% This process -- sometimes referred to as \textit{NN-Descent} -- empirically converges to a close approximation of the true graph in a small number of iterations~\cite{dong2011efficient}. 

% Given an approximate graph, we may use the same iterative procedure to \textit{refine} the neighbors of each node. 

% To construct an $\alpha$-NN graph, one may use LSH to obtain an initial graph

% \blue{we start the paragraph talking about both $k$-NN and $\alpha$-NN, but now we only mention $k$-NN (in this sentence and in the next paragraph)?}
% Good point: fixed to include alpha-NN (which is less common in practice but still used) 

In the past, high-performance algorithms such as SWG used the graph directly~\cite{malkov2014approximate}. 
However, the fastest indices no longer use the raw $k$-NN or $\alpha$-NN graphs. 
Instead, the final search index is obtained by removing unnecessary edges (or \textit{pruning}) and by adding edges (or \textit{diversifying}) in order to improve the navigability of the graph. 
Each graph index implements slightly different methods to accomplish this improved navigability, but they are all driven by similar principles inherited from \citet{arya1993approximate}.
Below, we describe several algorithms from the top of the ANN-benchmarks leaderboard.
% Our objective in this section is to argue that reordering is an orthogonal improvement to most graph-based indices, since the important graph properties remain the same.

\textbf{HNSW.}
Arguably the most popular graph index, HNSW is widely used for industrial applications~\cite{malkov2018efficient}. The HNSW index consists of a layered graph where each layer contains progressively more points from the dataset. Layer $l$ is a pruned $k$-NN graph of all nodes from layer $l-1$, with the addition of $O(e^l)$ new nodes not contained in previous layers. Because the layer populations increase exponentially, the upper layers contain relatively few nodes while the lowest layer is a $k$-NN graph of the complete data. The intuition behind HNSW is that we may examine the upper layers to quickly identify a good neighborhood for further exploration in the lowest level. HNSW directly uses the pruning heuristic from \citet{arya1993approximate}, which is as follows. Suppose that node $B$ is one of the $k$ nearest neighbors of node $A$. If node $B$ is closer to one of the other neighbors of $A$ than it is to $A$, then prune the link between $A$ and $B$.

% The intuition behind HNSW is that the $k$-NN graphs of the upper layers have long-distance links, since they are built on a highly downsampled version of the dataset. These links function as shortcut connections.
% The intuition behind HNSW is that the upper layers provide long-range shortcut connections while the lowest level is a pruned $k$-NN graph 
% \blue{I don't understand this intuition plus using levels here instead of layers before where they seem to refer to the same thing is confusing ... please fix}. 
% \blue{wouldn't these be the same as all the points in layer $l-1$? Why include all layers from $0$ since each one is fully contained in the next one?} 

% We traverse the upper levels of the graph via greedy search until we arrive at the bottom-most layer, which contains all points in the dataset. Here, we perform an exhaustive beam search to explore the neighborhood and locate the nearest neighbors. 

% The intuition behind PANNG is that some edges in a $k$-NN graph are redundant because there are likely alternative paths that connect the two nodes. 
% We may safely remove these edges, since graph search will simply take the other path. 

\textbf{PANNG and ONNG.} PANNG prunes a $k$-NN graph to remove edges when alternative paths exist between two nodes~\cite{iwasaki2016pruned}. 
The PANNG algorithm first adds edges to the $k$-NN graph so that the graph is fully connected and undirected. 
Then, the algorithm removes edges according to the following pruning heuristic. 
Given two nodes $A$ and $B$, if $B$ is not one of the top $k_r$ neighbors of $A$ and there exists a different path between $A$ and $B$ of length $\leq p$, then prune the edge. 
The values of $k_r$ and $p$ are hyperparameters. When $p = 2$ and $k_r = 0$, the PANNG heuristic prunes the same links as the HNSW heuristic, except under a corner case that is rare in practice\footnote{This occurs when the intermediary node $C$ on the path from $A$ to $B$ is not a neighbor of $B$. This is rare when $B$ and $C$ are near neighbors of $A$ because $d(B,C) \leq d(A,B) + d(A,C)$.}. ONNG extends PANNG with an additional diversification step to ensure that each node has in-degree at least $k_I$, then applies the same heuristic with $p=2$~\cite{iwasaki2018optimization}. In other words, ONNG ensures that all nodes are reachable, then removes the longest edge from all triangles in the graph.

\textbf{Other Algorithms.} Several other contributions, such as FANNG~\cite{harwood2016fanng} and methods based on NN-Descent, have at one point been strong contenders for the fastest search algorithm. The pruning techniques employed by such methods are similar to those described above. We refer the reader to \citet{shimomura2020survey} for recent review and thorough discussion.

\textbf{Similarities among Graph Indices.} 
Most graph algorithms begin by adding ``long distance'' edges between nodes and then pruning edges when a more desirable alternative path can be found. Under common hyperparameter choices, the pruning heuristics all behave similarly. It should come as no surprise that the resulting graphs have approximately the same performance on real-world search tasks~\cite{aumuller2020ann}. 
As previously mentioned, we expect these similarities to drive the generalization of our results (here illustrated for HNSW) to other popular graph indices based on the same principles.

% In practice, HNSW, ONNG, and PANNG . This is likely because the link pruning and diversification heuristics of each method have similar behavior. All three algorithms remove long ``shortcut'' paths between nodes that are already connected while maintaining edges that connect a node to its very nearest neighbors. With minor variations, the algorithms all use beam search over the pruned graph. The performance differences between graph kNN libraries seem to be from implementation details and C++ optimizations rather than major algorithmic innovations. In this paper, we implement and analyze HNSW graphs. However, we expect that our conclusions will also hold for other graphs because of the fundamental similarity between these algorithms.

\subsection{Searching Near Neighbor Graphs}
Graph-based near neighbor search consists of a walk along the edges of the graph. 
At each node $x$ encountered during the walk, we record the distance $d(q,x)$ between the query $q$ and the current node.
We explore the graph until we arrive at a node which is closer to the query than any candidate or neighbor of a candidate seen so far.
% \blue{But this is trivially true for the initialization ... is this stopping criterion valid after some period of exploration?}

\textbf{Beam Search.} 
There are two search algorithms in widespread use: greedy search and beam search\footnote{We use the definition of beam search from \citet{prokhorenkova2020graph}. The graph literature uses a slightly different definition, where every candidate is explored (rather than the best one).}. 
In greedy search, we simply choose the nearest node to the query from the neighbors of our current node. 
We stop the search process once none of the outgoing edges lead to a node that is closer than the current node. 
Beam search is conceptually similar to greedy search, but we maintain a dynamic list of $M$ candidates to investigate. 
Beam search allows us to explore outbound paths from any of the best $M$ nodes seen so far. 
If none of the outgoing edges of the $i$\textsuperscript{th} candidate lead to a closer node, then we consider the neighbors of candidate $(i+1)$. 
Once none of the $M$ nodes have edges that lead to closer neighbors, we return the best points from the list of candidates as our search results. 
Note that when $M = 1$, beam search and greedy search are equivalent.

% In this way, beam search explores paths through an additional $M$ nodes from the neighborhood of the point selected via greedy search. 
% Rather than exploring the neighbors of the best node at each iteration, we may select from the neighbors of the best $M$ nodes seen so far. 

\textbf{Beam Search Initialization.} 
Many recent innovations in graph $k$-NN reduce to \textit{initialization methods} that produce a good starting position for beam search. 
For example, HNSW uses greedy search over the hierarchical portion of the graph to find a seed node in the lowest layer, which is used to initialize beam search. 
PANNG and ONNG traverse a space-partitioning tree to extract seed nodes.
Alternative options include initialization via locality-sensitive hashing, node centrality measures, and even random selection. 
Recent experiments suggest that the type of initialization is immaterial to performance~\cite{lin2019comparative,lin2019graph}. 
\citet{prokhorenkova2020graph} provide theoretical support for this idea by proving that beam search over an $\alpha$-NN graph solves the approximate near neighbor problem given a ``sufficiently close'' initialization that can be found by repeated random sampling. \citet{laarhoven2017graph} prove a related result for greedy search, which relies on a similar initialization.

Initialization is also fairly inexpensive. 
To illustrate this, we constructed an HNSW graph on the SIFT1M dataset with $k = 8$ edges per node. 
We used beam search with $M = 500$ and found that the hierarchical part of the search process was responsible for only 2.9\% $(\pm1.9\%)$ of the query time. We then replaced the hierarchy with a process where we randomly select 50 nodes and use the closest option as the initialization, to reproduce the experiments by \citet{lin2019comparative}. We found no statistically significant difference in terms of recall or query time over $10$k queries. 
The practical takeaway is that under typical query conditions, initialization is an important but computationally insignificant part of the search process. 
Differences in the initialization method do not affect the validity of our experimental results.

% Since this process is shared among all graph-based indices, our experiments and analysis apply broadly to 
% The authors of \cite{prokhorenkova2020graph} provide theoretical support that justifies selecting the initial node as the best of many \textit{randomly chosen possibilities}. 
% However, recent work suggests that the hierarchy is unnecessary and that link pruning is the critical component of HNSW~\cite{lin2019comparative,lin2019graph}.

\section{Graph Reordering}
In this section, we introduce the graph reordering methods that we apply to near neighbor graphs with the goal of speeding up the search. 
Formally, graph reordering can be seen as constructing a labeling or indexing function $P: V \to \{1, \ldots, N\}$ that assigns each node $v \in V$ in a graph to a unique integer index (or label) between $1$ and $N$ following a pre-specified rule or in order to maximize some objective. 
Many formulations are possible, but generally we require $P$ to map connected nodes to similar (nearby) labels. 
The function $P$ is then used as the memory layout for the graph, with node $v$ assigned to memory location $P(v)$. 

\subsection{Techniques}

We consider three broad categories of reordering algorithms. 
The first category contains algorithms that optimize $P$ to maximize an objective function. 
The optimization problem is typically NP-hard, so we are usually forced to accept an approximate solution for $P$. 
The second category assigns labels to nodes based on their degree or other local graph features. 
Finally, graph partitioning algorithms can be repurposed to work as reordering algorithms by assigning a contiguous memory range to each partition. 
Below, we review several recent algorithms from each category. 
These are considered in Section~\ref{S:experiments} to assess their ability to accelerate near neighbor search.

% \blue{I changed the description below quite a bit to make it clearer. Please check that the meaning is still correct}

\textbf{Gorder.} Gorder (graph-order) is a graph reordering algorithm that seeks to maximize the overlap between the neighborhoods of nodes with consecutive labels~\cite{wei2016speedup}. 
Indeed, Gorder finds $P$ by maximizing the number of shared edges among size-$w$ blocks of consecutive nodes. 
This is a good proxy for cache efficiency because a block that contains many overlapping nodes is likely to avoid a cache miss, as each node's neighbors are stored no further than $w$ memory locations away. 
Formally, the Gorder labeling function $P_{\mathrm{GO}}$ can be found through the following maximization
\begin{equation}\label{E:Gorder}
P_{\mathrm{GO}} = \underset{P}{\mathrm{arg\,max}} \!\!  {\sum_{\substack{u,v \, \in \, V \mathrm{s.t.} \\|P(u) - P(v)| < w}}} \!\! S_s(u,v) + S_n(u,v),
\end{equation}
where $S_s(u,v)$ indicates whether $u$ and $v$ are directly connected and $S_n(u,v)$ counts how many common neighbors they have.
In other words, Gorder maximizes the \textit{average neighborhood overlap} between nodes that are within $w$ positions of each other under the labeling function. 
Intuitively, two nodes will be placed together if they share a direct edge or, even if that is not the case, if they share many common neighbors.
Maximizing the objective in~\eqref{E:Gorder} is NP-hard, but the Gorder algorithm in~\citet{wei2016speedup} provides a $1/(2w)$-approximate solution.

\textbf{Reverse Cuthill Mckee.} 
The \textit{bandwidth} of a matrix is defined as the maximum distance of a nonzero element from the main diagonal. 
The Reverse Cuthill Mckee (RCM) algorithm~\cite{cuthill1969reducing,george1971computer} is a reordering method originally introduced to minimize the bandwidth of a sparse symmetric matrix. 
If we apply this method to an adjacency matrix, then the corresponding optimization problem has an immediate interpretation in terms of graph reordering~\cite{auroux2015reordering}
\begin{equation}\label{E:RCM}
P_{\mathrm{RCM}} = \underset{P}{\mathrm{arg\,min}} \,\, \max_{(u,v) \in E} |P(u) - P(v)|,
\end{equation}
where $E$ is the edge set of the graph of interest.
From~\eqref{E:RCM} it follows that the RCM objective is to minimize the \textit{maximum label difference} between connected nodes.
Similar to Gorder, the problem is difficult: the exact solution is NP-complete. The RCM algorithm is a heuristic method based on breadth-first search to find a function $P$ with low (but not necessarily optimal) bandwidth.

\textbf{MLOGA and MLINA.} 
% \green{
The objective in~\eqref{E:RCM} motivated extensions that focus on an aggregated measure (as opposed to the maximum) of the discrepancies between labels of connected nodes.
Both the minimum logarithmic arrangement (MLOGA) and the minimum linear arrangement (MLINA), which originally arose in the context of social network compression, are examples of these approaches~\cite{chierichetti2009compressing}. 
More precisely, MLINA seeks to minimize the sum of label discrepancies
\begin{equation}\label{E:MLINA}
P_{\mathrm{MLN}} = \underset{P}{\mathrm{arg\,min}} \sum_{(u,v) \in E} |P(u) - P(v)|,
\end{equation}
whereas MLOGA first applies a logarithmic transformation to these discrepancies
\begin{equation}\label{E:MLOGA}
P_{\mathrm{MLG}} = \underset{P}{\mathrm{arg\,min}} \sum_{(u,v) \in E} \log (|P(u) - P(v)|),
\end{equation}
where, again, $E$ is the edge set of the graph of interest.
As expected, both problems are NP-hard. However, specialized heuristics have been developed to approximate the solution of both MLOGA~\cite{chierichetti2009compressing,safro2011multiscale} and MLINA~\cite{wei2016speedup}. 
% }

\textbf{Degree Sorting.} 
Degree sorting is a lightweight reordering algorithm based on the idea that high-degree nodes are likely to share many edges. 
Indeed, many practical graphs obey a power-law degree distribution, where a small number of nodes form a densely connected sub-graph. %share many connections. 
For undirected graphs with a power law degree distribution, degree sorting will create a group of neighboring high-degree nodes in the first contiguous memory block. 
To obtain $P$, we simply sort the nodes in descending degree order and let $P(v)$ be the sorted rank of node $v$. 
Since degree sorting only requires local node information, it is orders of magnitude faster than the optimization-based methods in~\eqref{E:Gorder}-\eqref{E:MLOGA}.

\begin{figure*}[t]
\centering
\mbox{\centering
\includegraphics[width=2.2in]{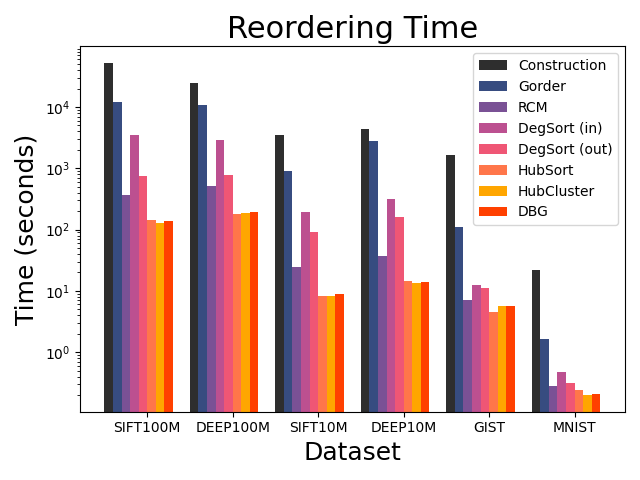}
\includegraphics[width=2.2in]{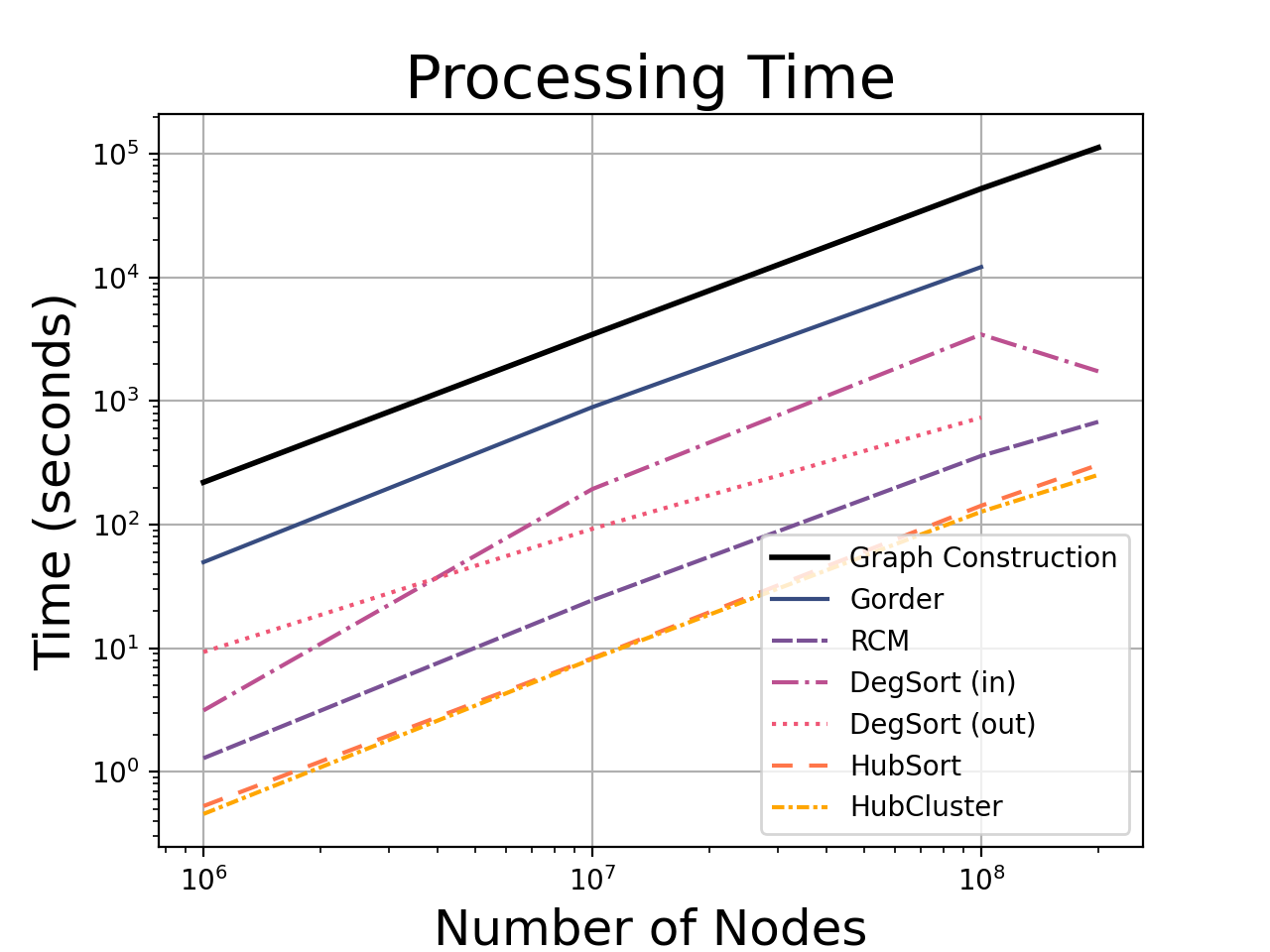}
\includegraphics[width=2.2in]{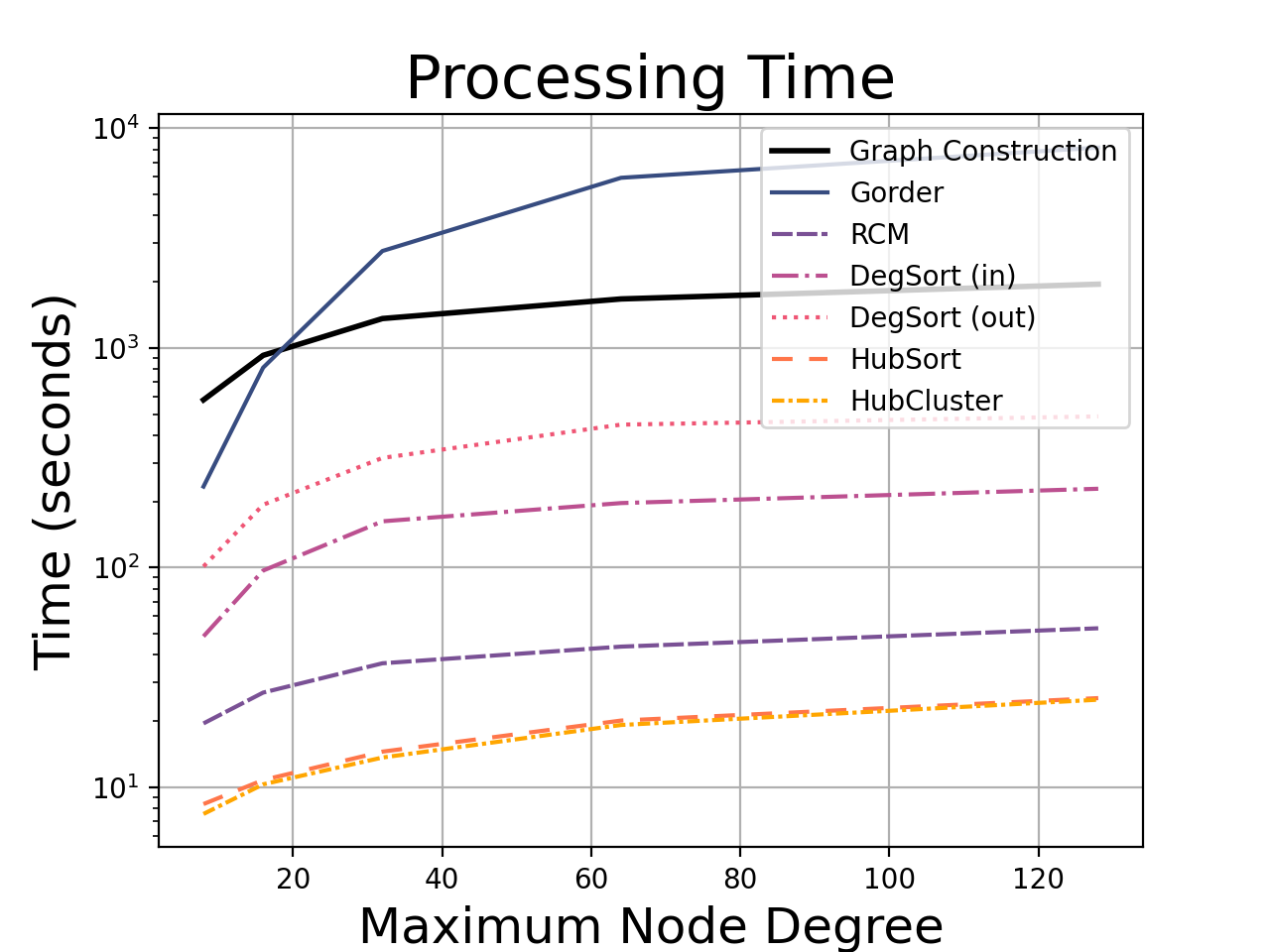}
}
\vspace{-0.1in}
\caption{Graph reordering scales to large datasets and complex near neighbor graphs. In our experiments, the reordering algorithms required less time to run than graph construction (left). Reordering is feasible even for large graphs with many nodes (SIFT with $k = 16$, middle) and for densely connected graphs ($k \leq 120$) on large datasets (GIST1M, right).}
\label{fig:profileplot}
\end{figure*}

% Why is graph reordering appropriate for NNS?

\begin{figure}[t]
\centering
\mbox{\centering
\includegraphics[width=2.5in]{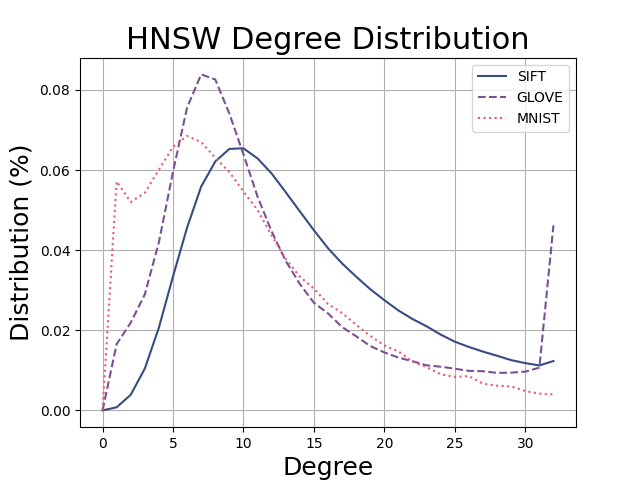}}
\vspace{-0.1in}
\caption{In practice, pruned $k$-NN graphs are far from being $k$-regular. The figure shows the empirical out-degree distribution of the HNSW base layer of three graph indices. Although we begin with $k = 32$ outbound edges per node, the pruning heuristic removes over half of the edges from the network.}
\label{fig:degreedist}
\end{figure}

\textbf{Hub Sorting.} 
Hub sorting~\cite{zhang2017making} is similar to degree sorting, but with the caveat that we only sort hubs (nodes with many connections). 
The algorithm first splits the nodes into two groups (hubs and non-hubs) based on a degree threshold. 
The threshold is set as the average degree, so that hubs are defined as nodes with greater-than-average degree. 
To find $P$, the hubs are sorted by degree and the non-hubs keep their original ordering. 
Experiments have show that degree and hub sorting are not always beneficial: they may even slow down graph processing if the original ordering of the graph had good locality properties~\cite{balaji2018graph}. 
\textit{Selective} hub sorting was introduced to address this issue. 
In selective hub sorting, we only reorder a graph if the hubs are densely connected to each other. 
To decide whether to sort the graph, we compute the packing factor, a computationally inexpensive diagnostic value that predicts whether hub sorting is likely to be effective~\cite{balaji2018graph}.

\textbf{Hub Clustering.} 
Hub clustering is the same as hub sorting, but without sorting the hubs after separating high and low degree nodes~\cite{balaji2018graph}. 
The primary advantage of hub clustering is that it is an incredibly lightweight reordering algorithm that can be accomplished in a single pass through the graph.

\textbf{Degree-Based Grouping.} 
Degree-Based Grouping (DBG) \cite{faldu2019closer} is an extension of hub clustering to multiple groups. 
Nodes are divided into $w$ groups based on degree ranges associated with each group. 
The groups are sorted in descending degree order, but the order of the nodes within each group is not changed. 
The authors of~\cite{faldu2019closer} use logarithmically spaced thresholds to evenly distribute nodes among groups for power-law graphs.
Since k-NN graphs do not have power-law degree distributions, our implementation uses the $w$ quantiles of the degree distribution instead.

\textbf{Graph Partitioning.} 
Graph partitioning is a very well-studied area with a large number of established methods. 
One can create a graph reordering algorithm from any graph partitioning method by using the graph partitioning method to assign nodes to groups and using the same labeling scheme as in DBG.

\subsection{Greedy Search and Graph Reordering}

In this section, we show that objective-based graph ordering methods improve the complexity of greedy search under the idealized cache model of computation.

% \subsection{Cache Performance of Beam Search}

% Insert plot of degree distribution of

\section{Experiments}
\label{S:experiments}

\textbf{Experiment Setup.} 
Our experiments measure index performance using wall-clock query time under different reordering strategies. 
Since cache misses are sensitive to a variety of external factors, we took extra care to design an objective comparison. 
For example, multicore systems often share the L3 CPU cache, allowing cache performance to be affected by unrelated programs running on a different core. 

To obtain accurate and repeatable results, we designed our experiments to mimic real-world production environments while controlling as many external factors as possible. 
We benchmark the system after a ``warm start,'' where assets such as the index, query and, data are pre-loaded into RAM. 
To avoid the difficulties associated with timing very short events, we record the total time needed to perform $10$k queries and report the average query time. 
Finally, we restrict the query program to a single core, and we ensure that the benchmark is the only program running on the server. 
We run all of our experiments on a server with 252 GB of RAM, 28 Intel Xeon E5-2697 CPUs, and a shared 36 MB L3 cache. 

We measure cache miss rates using the Linux perf tool to record hardware CPU counters.
The perf tool measures hardware counters for events such as data reads and cache hits. We use these counters to compute the cache miss rate by dividing the number of data cache misses by the number of data references.
We run our most successful reordering methods (Gorder and RCM) on SIFT100M and report results for the L1, L2, L3 and TLB caches. 
To validate these measurements, we also provide experiments with the cachegrind tool for all reordering algorithms. 
Cachegrind runs the program through a virtual processor that records every instruction and cache reference while executing the program. 
We annotate the source code of our HNSW program and report the cache misses for the lines of code which execute node traversals and distance computations. 

% Since it is very computationally expensive to benchmark HNSW with cachegrind, we only run the tool for the SIFT100M dataset. 

\begin{table}
\caption{Perf stat results for cache misses and TLB misses on SIFT100M. We report the L1, L2, L3 misses as the hardware counter for misses divided by the total number of data loads.}
\vspace{2mm}
  \centering
  \begin{tabular}{ l c c c c} 
\toprule
Algorithm & L1 (\%) & L2 (\%) & L3 (\%) & TLB (\%) \\
\midrule
Original & 19.53 & 13.9 & 6.5 & 3.85\\
RCM & 17.37 & \textbf{7.61} & 5.1 & 2.56 \\
Gorder & \textbf{14.46} & 9.6 & \textbf{4.0} & \textbf{2.14} \\
\bottomrule
\end{tabular}
\vspace{-0.2cm}
\label{tab:perfstat}
\vspace{-0.2cm}
\end{table}

% We considered using a sampling tool such as perf, but an instrumentation profiler is required to compare the absolute numbers of cache misses between reordering algorithms. 
% To measure cache miss rates, we use the perf tool to read CPU performance counters. CPU performance counters are hardware registers that record the number of events such as cache misses, instruction executions and other low-level hardware events that occur during the execution of a program. To compute the cache miss rate, we divide the number of data cache misses by the total number of data cache load operations.

% To measure the L1 cache miss rate, we use the Linux perf tool to record CPU performance counters for an application running. To measure lower-level cache miss rates, we emulate the CPU using cachegrind and report the L2 and LL miss rates. 
\begin{figure}[b]
\centering
\vspace{-0.4cm}
\includegraphics[width=2.5in]{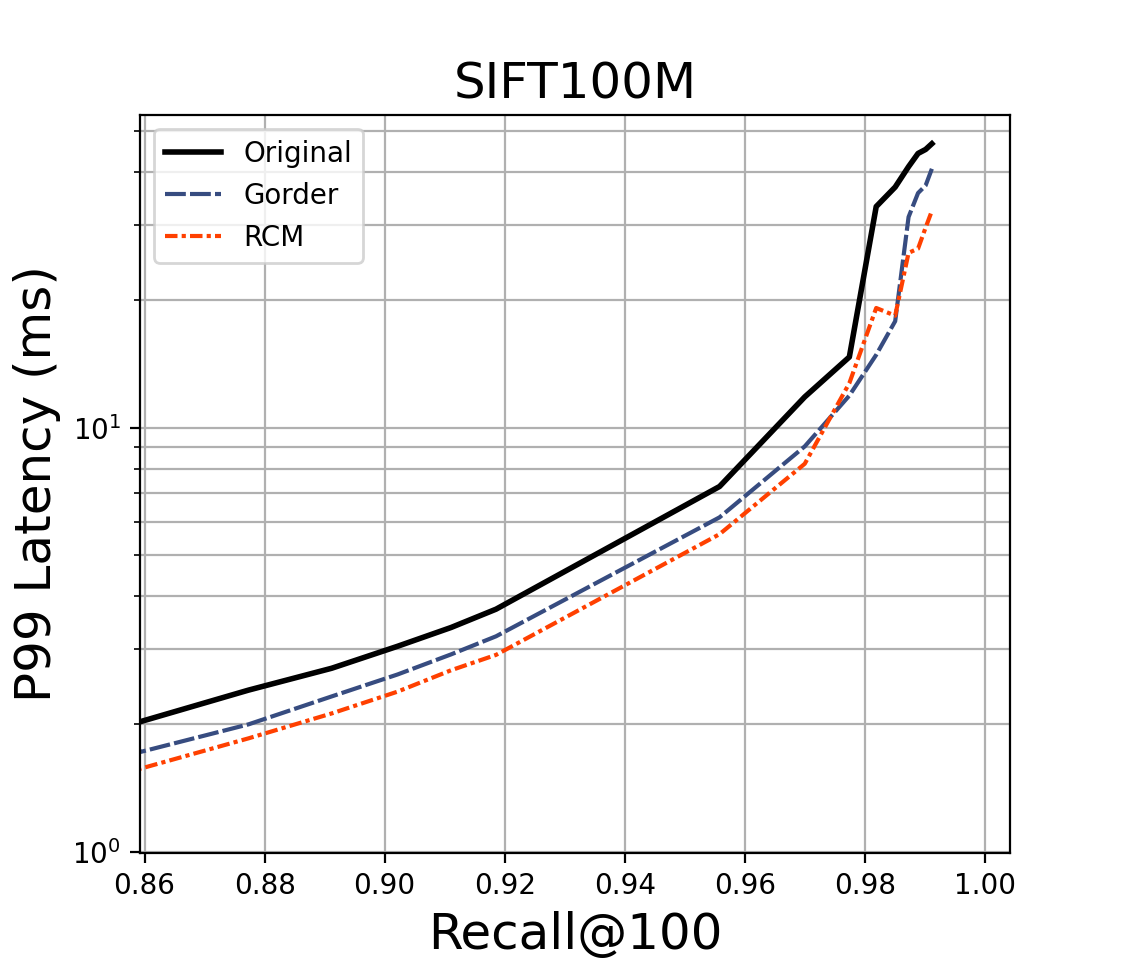}
\caption{99\textsuperscript{th} percentile of latency (P99) for SIFT100M (lower is better). Note the log scale. We observe an improvement of 17\% with RCM and 30\% with Gorder.}
% \vspace{-1cm}
\label{fig:p99}
\end{figure}

\begin{figure*}[t]
\centering
\mbox{\centering
\includegraphics[width=2.2in]{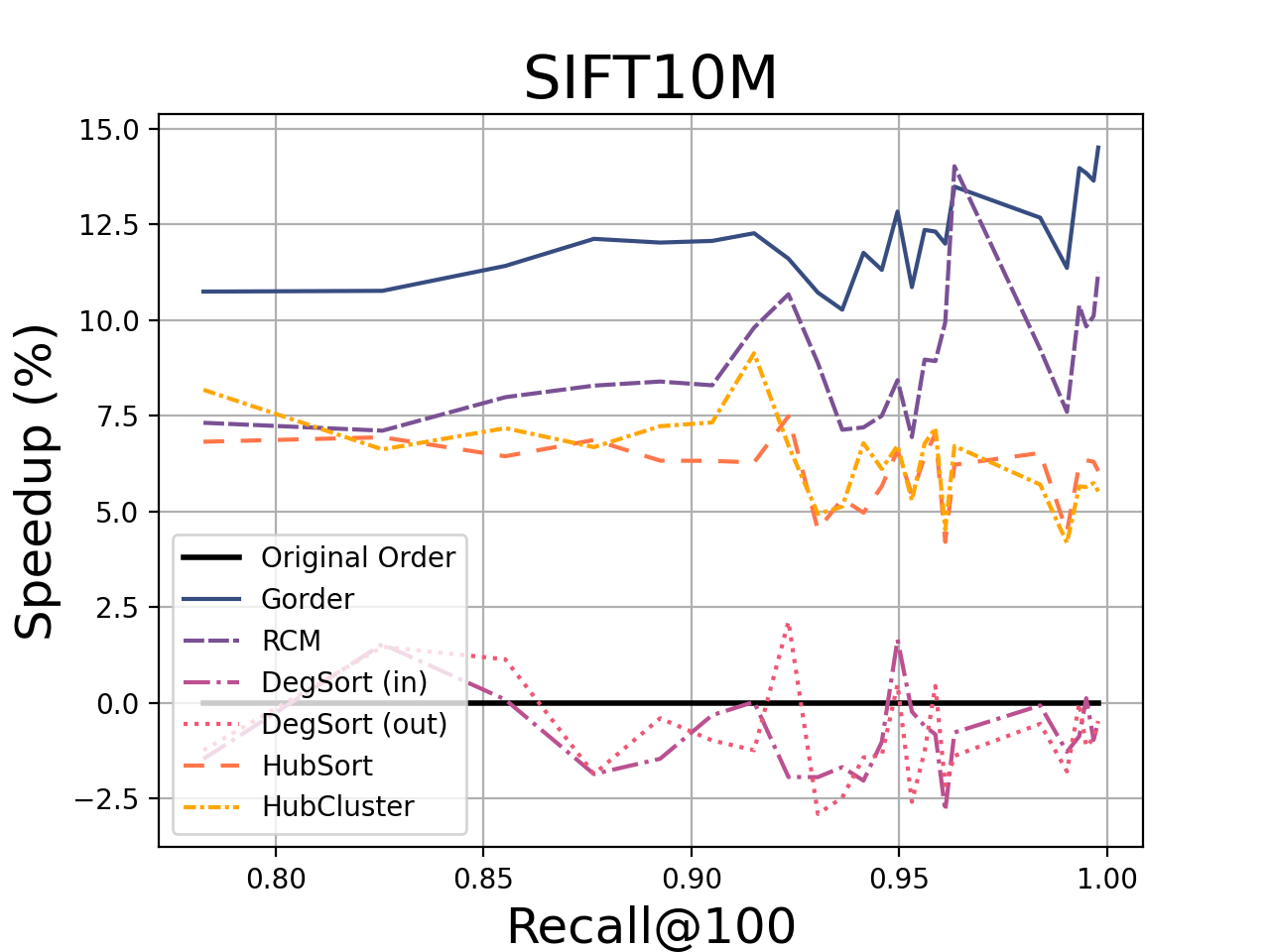}
\includegraphics[width=2.2in]{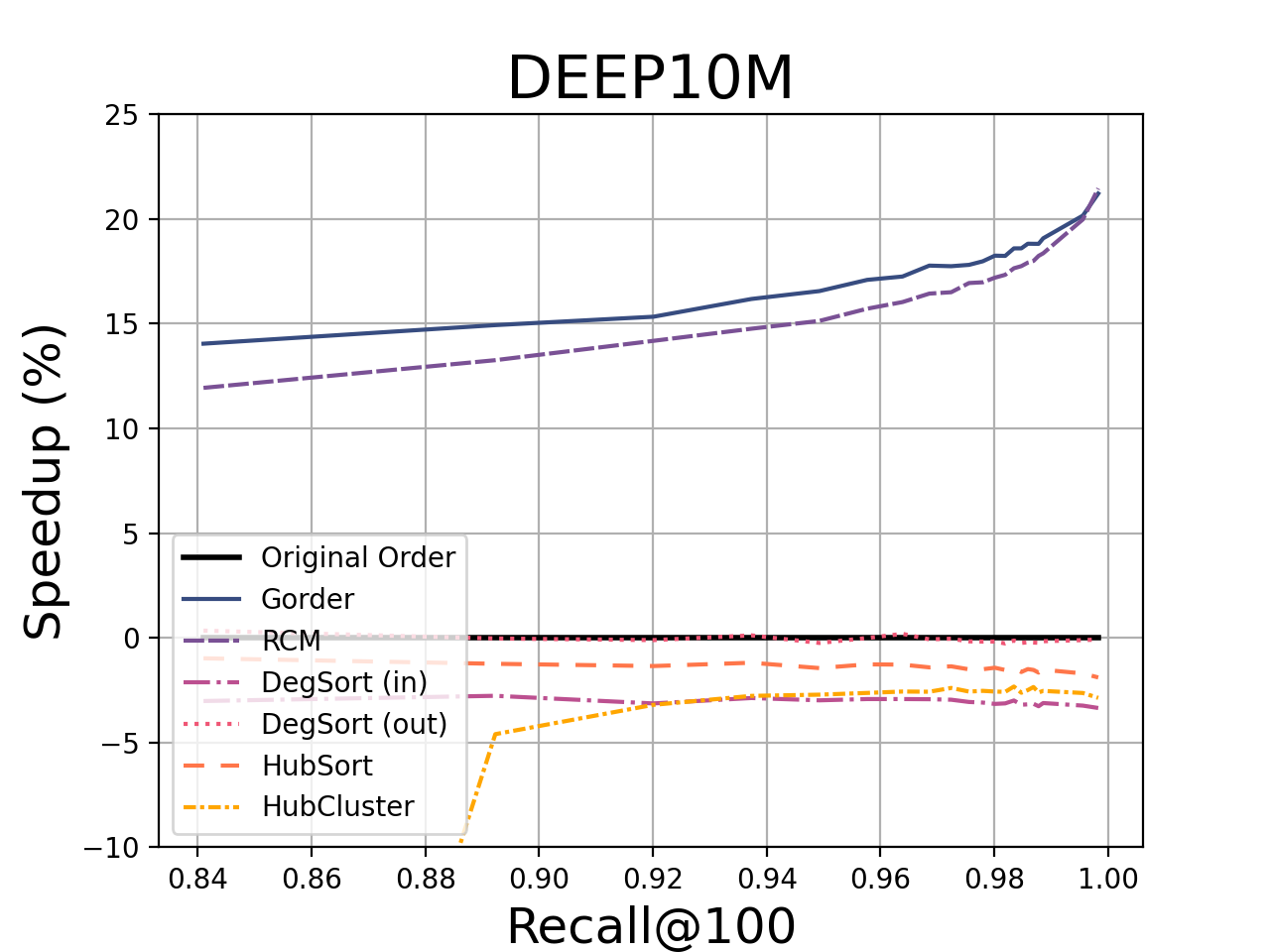}
\includegraphics[width=2.2in]{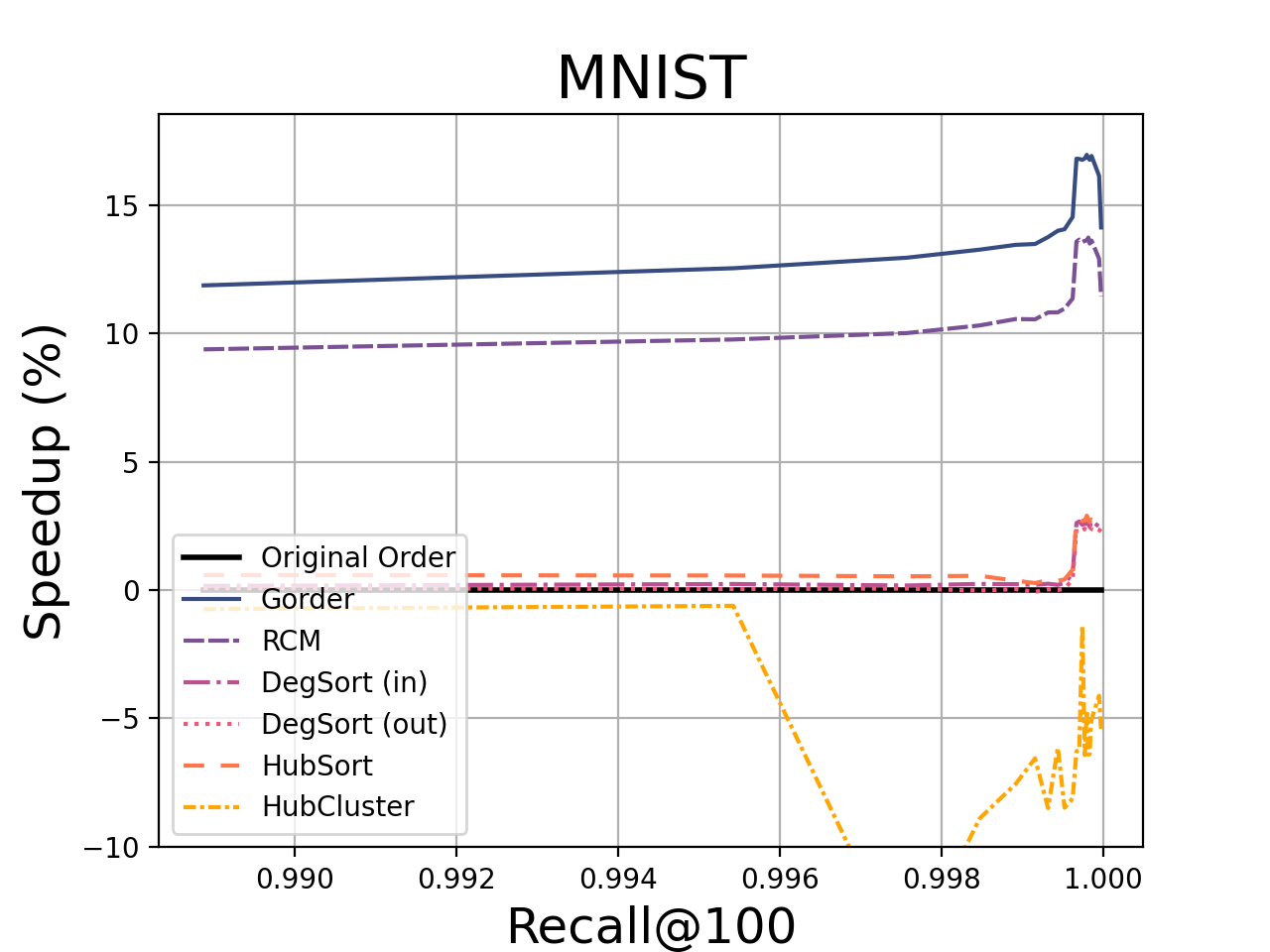}
}
\mbox{\centering
\includegraphics[width=2.2in]{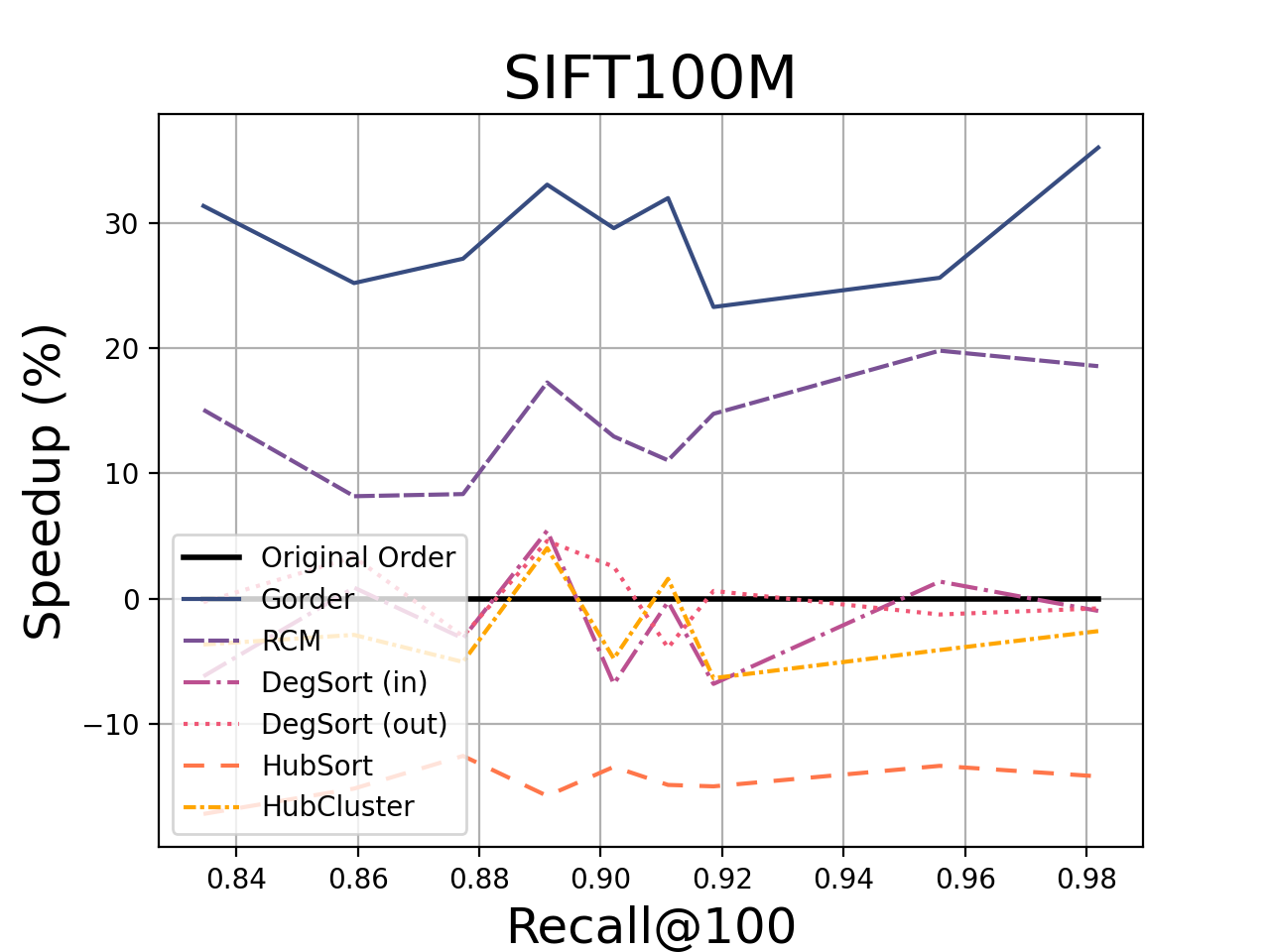}
\includegraphics[width=2.2in]{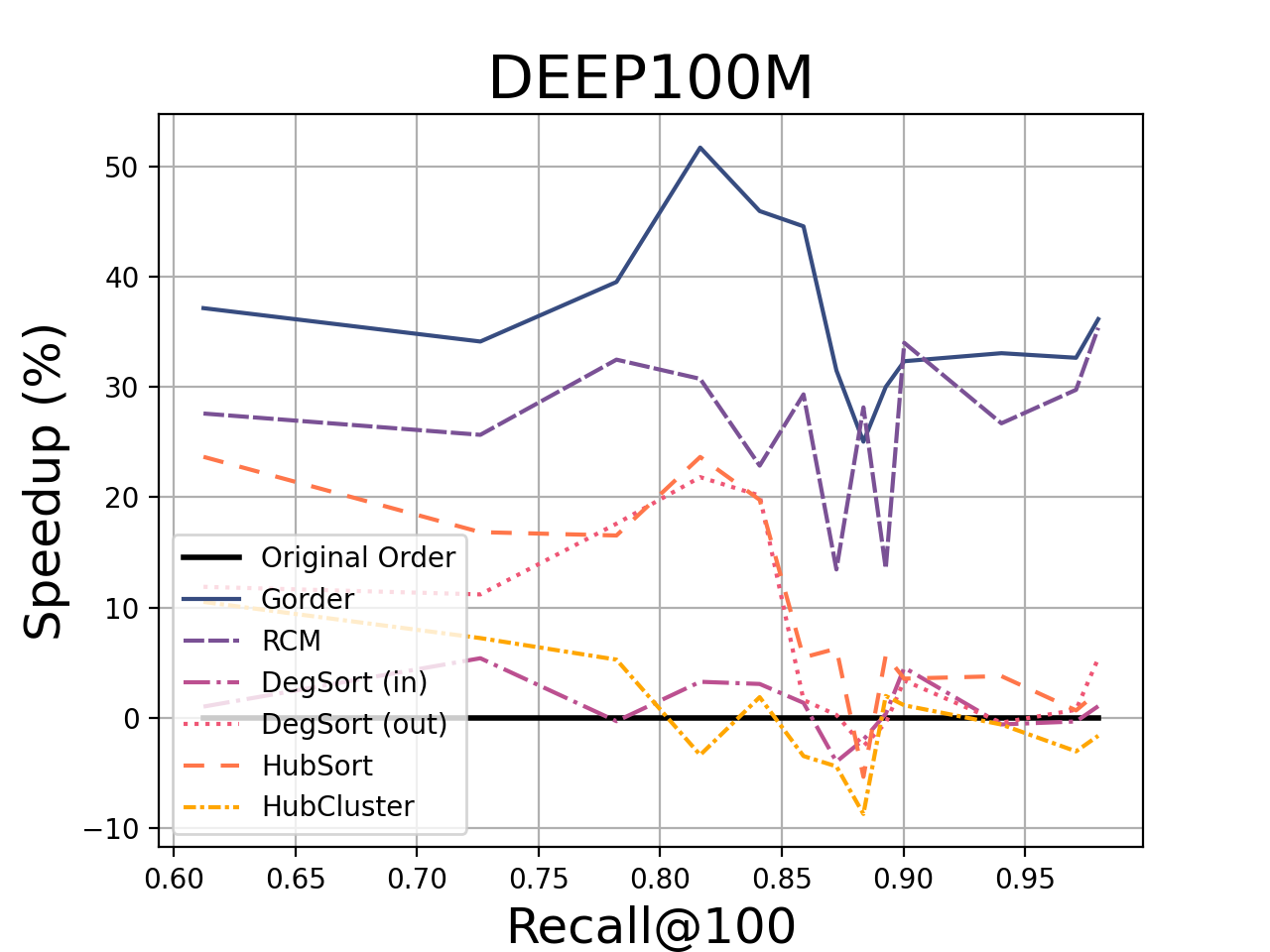}
\includegraphics[width=2.2in]{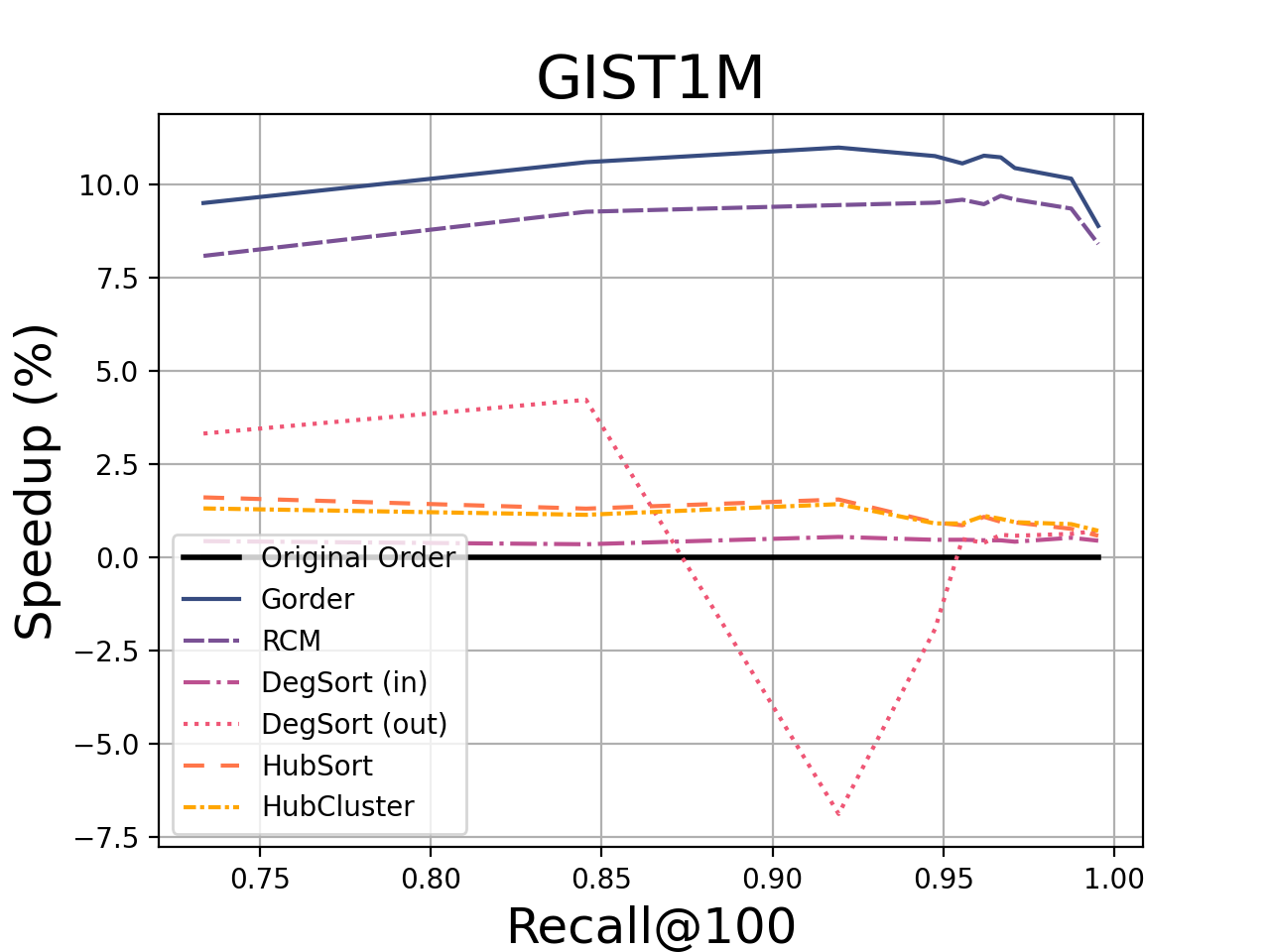}
}
\vspace{-0.1in}
\caption{Effect of graph reordering on the query time vs. recall tradeoff. Reordering algorithms above the black line have faster query time than the original ordering. Algorithms below the line cause a slowdown. The speedup changes with the recall because the memory access pattern of beam search changes with the buffer size.}
\label{fig:accuracyplot}
\end{figure*}

\begin{table*}

  \centering
  \addtolength{\tabcolsep}{-4pt} 
  \begin{tabular}{ l|c|c|c|c|c|c|c|c } 
\toprule
 &
Original &
Gorder &
RCM &
DegSort (in) &
DegSort (out) &
HubSort &
HubCluster &
DBG\\
\midrule
L1 (\%) & 22.76  & 19.28  &  20.61 & 22.76  & 22.76  &  22.85  & 22.77  & 22.81 \\
L3 (\%) & 13.56  &  8.32  & 8.91  & 13.55  &  13.56  &  13.63 & 13.51  & 13.34 \\
\bottomrule
\end{tabular}
% \addtolength{\tabcolsep}{4pt}
\vspace{-0.2cm}
\caption{Cache miss rates for node traversals and distance computations on SIFT100M (lower is better). It should be noted that due to the 100x latency difference between cache and RAM, small improvements to the cache miss rate can substantially speed up an algorithm. The ranking of algorithms by cache miss rate agrees with the ranking of algorithms by speedup in Figure~\ref{fig:accuracyplot}.
}
\label{tab:cachemiss}
\vspace{-0.2cm}
\end{table*}

\textbf{Implementation Details.} 
We extended the nmslib implementation of HNSW to support graph reordering. 
We implemented minor changes to the index to speed up graph construction and facilitate reordering. % For example, we pre-allocate memory for edges and nodes rather than dynamically allocate the index during construction. 
However, the memory access pattern and computational aspects remain unchanged. 
To ensure that our performance numbers are realistic, we verified that our implementation uses the same memory layout, requires the same number of distance computations, and produces the same graph as nmslib-HNSW.
We use the optimized, flat layout for nodes to avoid memory fragmentation and provide the most competitive baseline possible. We represent links as a fixed-size array of integers and embed the data alongside the node label and link array.
% use the same memory layout as the nmslib-HNSW library. We also verified that our implementation requires the same number of distance computations, produces the same graph, and has the same query time as the nmslib implementation.

\textbf{Objective-Based Reordering.} 
Since our $k$-NN search benchmarks involve datasets with millions of nodes, we avoid algorithms that are known to scale poorly to large graphs. We implemented Gorder and RCM, but we did not implement MLOGA or MLINA because these algorithms have very high runtime without tangible improvements over Gorder or RCM~\cite{wei2016speedup}.

% achieve state-of-the-art \red{reordering performance} \blue{What does reordering performance mean?}. 
% The benchmarks in~\citet{wei2016speedup} show that Gorder and RCM are the two \red{best-performing methods} \blue{again, best-performing in what? because that paper is not about NN search, right?}. 

% We also did not implement graph partitioning algorithms for the same reason. 
% Prior work shows that Gorder and RCM beat graph partitioning in terms of both reordering time and \red{speedup} \blue{speedup of what? Please clarify this because it might seem that you are talking about NN speedup when previously we said that no one has looked into this before us} after reordering. 

\textbf{Degree-Based Reordering.} 
We follow the suggestion of~\citet{balaji2018graph} to use the \emph{in-degree} as a local feature for hub clustering and degree-based methods because beam search is a so-called push application \footnote{See Section 2 of ~\citet{balaji2018graph} for the definition of push/pull applications.}. We use the average degree as the threshold for hub clustering and sorting, and we use 8 groups given by the 8 quantiles of the degree distribution for DBG.

At first glance, it may seem that degree-based reordering algorithms are inappropriate for near neighbor search because the ideal $k$-NN graph has a constant degree distribution. 
However, this is not true. 
Popular graph algorithms accept the \textit{maximum} number of links as a hyperparameter, but the pruning and diversification heuristics produce graphs with nontrivial variations in node degree. 
Figure~\ref{fig:degreedist} shows the out-degree distribution of several real-world near neighbor graphs. % when the maximum degree is set to 32. %\blue{are these in or out degrees?} 
While these graphs do not follow a true power-law distribution, there is enough variation that we may reasonably expect lightweight reordering to perform well. 

\textbf{Datasets.} 
We use large datasets that are representative of embedding search tasks. 
We perform experiments on datasets from ANN-benchmarks as well as on 10M and 100M sized subsets of the SIFT1B and DEEP1B benchmark tasks. 
Table~\ref{tab:datasets} contains information about our datasets. 
In all of our experiments, we request the top 100 neighbors of a query and we report the recall of the top 100 ground truth neighbors.

\begin{table}
\caption{Datasets. Each dataset has $N$ entries, $d$ features, and requires ``vector size'' bytes for each entry.
}
\vspace{2mm}
  \centering
  \begin{tabular}{ l c c c } 
\toprule
Dataset & $N$ & $d$ & vector size \\
\midrule
GIST & 1 M & 960 & 3.8 kB\\
SIFT & 10 - 100 M & 128 & 128 B\\
DEEP & 10 - 100 M & 96 & 384 B \\
MNIST & 60 k & 784 & 3.1 kB \\
\bottomrule
\end{tabular}
\vspace{-0.2cm}
\label{tab:datasets}
\vspace{-0.2cm}
\end{table}

\textbf{Hyperparameters.} 
For construction, HNSW requires the determination of two parameters: the maximum number of edges for each node ($k_c$) and  the size of the beam search buffer ($M_c$) used to find the $k_c$ neighbors during graph construction. 
We set $M_c = 100$ and construct indices for $k_c \in \{4, 8, 16, 32, 64, 96\}$. 
From these options, we select the index with the best recall-latency trade-off (for recall $> 0.95$) and use that index for our graph reordering experiments. 
While it is true that the best hyperparameters change based on the recall, there is typically a clear winner for the high-recall regime. 
To query the index, HNSW requires the beam search buffer size ($M_q$), which controls the trade-off between recall and query latency. 
We reorder the graph and issue the same set of $10$K queries for each ordering. 
When we query the index, we vary the beam search buffer size $M_q$ from 100 to 5000.

\textbf{Results.} 
Figure~\ref{fig:accuracyplot} shows the effect of graph reordering on the query time vs. recall tradeoff. We report the R100@100 recall, or the recall of the top 100 ground truth neighbors among the top 100 returned search results.
For a given recall value, we calculate the speedup as the ratio of average latency without/with reordering, where the averages are over 10K queries and 5 runs.
Figure~\ref{fig:profileplot} presents the cost of reordering the graph using various methods and shows how reordering time scales with the number of nodes $N$ and the maximum degree $k_c$ of each node. 
Table~\ref{tab:perfstat} contains perf cache measurements and Table~\ref{tab:cachemiss} contains cachegrind information for graph search over the SIFT100M dataset. Because these two experiments have similar results, we may reasonably conclude that Gorder and RCM improve latency by reducing cache miss rate.
Finally, we measured the 99\textsuperscript{th} percentile of latency (P99) for the SIFT100M dataset in Figure~\ref{fig:p99}, to ensure that our altered graph layout does not increase the tail latency. We find that Gorder and RCM improve both the P99 and average latencies by at least 20\% in the high-recall regime.

\section{Discussion}

Our experiments suggest that graph reordering could become a standard preprocessing step to improve the query time, since it does not substantially inflate the index construction time. 
Reordering only affects the representation of nodes in memory and does not change the recall, search algorithm, or other properties of the graph index. %\red{harm the index} \blue{not sure what you mean by this. The index is the same, right? we might take longer or shorter to traverse it, but the HNSW graph index that we generate does not depend on the reordering technique chosen, is this correct?}. 
Objective-based reordering algorithms are most effective for this problem, with a typical speedup of 10\% on small datasets ($N < 1$M) and speedups of up to 40\% on large datasets and in the high-recall regime (Figure~\ref{fig:accuracyplot}). 
The Gorder and RCM algorithms consistently yield speedups in all our experiments.
This likely occurs because the reordering objective is a good proxy for cache coherence. It is reasonable to conclude that objective-driven methods will rarely (if ever) slow down the index.
However, lightweight reordering algorithms are far less effective for near neighbor search than for applications previously considered in the literature. Although degree-based clustering does perform well on some tasks (e.g. SIFT10M), near neighbor graphs seem to have a pathological degree distribution for such methods. % that rely on local node features.

\textbf{Cost of Reordering.} 
Most studies of graph reordering are focused on applications such as PageRank, where the graph is processed a small number of times to obtain an output. 
In such cases, the objective is to obtain an \textit{end-to-end} speedup, which penalizes long graph reordering times. 
Such applications favor lightweight (but less effective) algorithms over more complex reordering techniques. 
However, the production requirements of near neighbor search are exactly the opposite: a single index may be queried millions of times over its life cycle. 
Near neighbor graph construction is already an expensive offline task, so the runtime disadvantage of graph reordering is less problematic for near neighbor search than for other applications. 

Nonetheless, we observe that the graph reordering time is, in many cases, negligible when compared to the graph construction time. 
For example, our most expensive reordering algorithm (Gorder) was a factor of 10x faster than HNSW construction for most datasets (Figure~\ref{fig:profileplot}). 
Even though the reordering time of Gorder scales quadratically with degree, reordering is still feasible even for large graph indices on datasets such as DEEP10M with many connected neighbors (up to $k = 120$). % \blue{This might be confusing since the figure does not show the x-axis all the way up to 200}
This is likely a consequence of near neighbor graph construction heuristics, which decrease the average node degree with aggressive pruning. 

\textbf{Benefits of Reordering.} 
Reordering is an index-agnostic method to improve the performance of graph-based near neighbor search. 
Because reordering depends on the node access pattern of beam search, which is common to most algorithms, reordering is applicable to a wide range of practical search tasks. 
It should be noted that most real-world deployments function under strict latency requirements: a maximum search time of 20 ms is a common constraint~\cite{nigam2019semantic}. 
A 20\% improvement in search time allows the system to perform more sophisticated processing of the search results, or alternatively to operate at a higher recall.

\textbf{Extensions.} Reordering is likely to exhibit synergy with other algorithms and performance tricks used in production systems.
Ideas from graph reordering may benefit partition-based search because recent algorithms for locality-sensitive hashing (LSH) use $k$-NN graph cuts to form data partitions~\cite{dong2020learning}. 
For example, one could reorder partition locations in memory to speed up multi-probe methods that access multiple nearby partitions.

Near neighbor graphs are also frequently combined with sample compression methods such as product quantization or other codebooks~\cite{jegou2010product}. 
Recent experiments with preprocessing transformations suggest that it is beneficial to perform the graph search on a subspace of reduced dimension~\cite{prokhorenkova2020graph}. 
This could amplify the effects of graph reordering.
When the data size is small, a larger sub-graph can fit into the CPU cache, improving the benefits of memory locality for graph processing. 
This idea is supported by our experiments, where we observe larger speedups for SIFT and DEEP than for MNIST and GIST, which have a large per-sample storage cost.
Thus, graph reordering may show even greater benefits when integrated with quantized search indices and dimensionality reduction.

\section{Conclusion}

We introduce graph reordering to the popular and important task of graph-based near neighbor search. We find that by relabeling the graph, we can obtain query-time speedups of up to 40\%. The reordering process is inexpensive when compared to graph construction, and our results apply broadly to most graph-based search indices. Therefore, we expect that reordering will become a standard preprocessing step for graph-based near neighbor search. 

\bibliography{main}
\bibliographystyle{icml2021}

\end{document}